\documentclass{aa}
\usepackage{graphics}
\begin{document}

   \title{ RX~J1643.7+3402 : a new bright cataclysmic variable
\thanks{ \small Partly based on observations obtained with the Byurakan
Observatory 2.6-m telescope and with the OHP 1.93-m and 1.2-m telescopes.}}

   \subtitle{}

   \author{A.M. Mickaelian \inst{1}, S.K. Balayan \inst{1},
           S.A. Ilovaisky \inst{2}, C. Chevalier \inst{2}, M.-P. V\'eron-Cetty
\inst{2}, P. V\'eron \inst{2}}
          \institute{ Byurakan Astrophysical Observatory and 
          Isaac Newton Institute of Chile, Armenian branch, Byurakan 378433, Armenia \\
\email{aregmick, sbalayan@bao.sci.am} \\
             \and Observatoire de Haute-Provence, CNRS, F-04870 Saint-Michel
l'Observatoire, France\\
\email{ilovaisky, chevalier, mira, veron@obs-hp.fr}
}

   \offprints{P. V\'eron}

   \date{Received 18 July 2001; accepted }

\abstract{We report the discovery of a new bright (V$\sim$12.6) cataclysmic variable star identified with the ROSAT X-ray source RX~J1643.7+3402. Spectroscopic and photometric observations show it to be a novalike variable sharing some of the characteristics of the SW Sex sub-class of novalike CVs. The spectroscopic period may be either 2\fh575 or 2\fh885, within the period ``gap." A photometric modulation with a probable period of 2\fh595 and an amplitude of $\sim$ 0.1 mag in V is present on most nights and could be either a ``positive" or a ``negative" superhump modulation (depending on the exact spectroscopic period), indicating the presence of a precessing accretion disk in this system. Rapid variations of 0.1 to 0.2 mag amplitude in V repeat with a time scale of $\sim$ 15 min.
  }
\authorrunning{Mickaelian et al. }

   \maketitle
\today

%

\begin{figure}[t]
\resizebox{8.8cm}{!}{\includegraphics{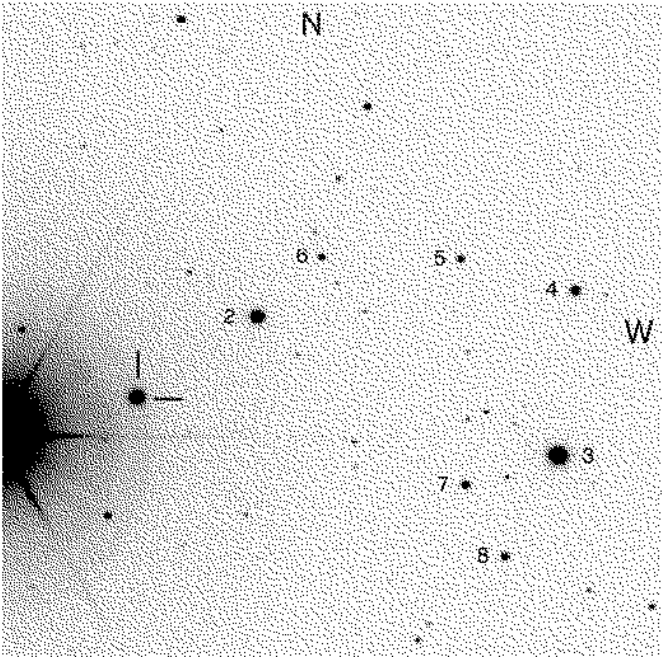}}
\caption{\label{figureA0}
Finding chart for RX~J1643.7+3402. This is a 30-sec R$_{\rm c}$ exposure
taken on September 23, 2000 with the OHP 1.2-m telescope.
The field of view is 6$\times$6 arc-min. The variable is
indicated by two tick marks. Magnitudes and colors for numbered field stars
are given in Table \ref{table2}. Stars 2 and 3 were used as comparisons. The bright star
(V=6) at left is HD~151087.}
\end{figure}

\section{Introduction}
In order to study the surface density of bright QSOs (Mickaelian et al. \cite{mickaelian}), we have conducted a search for such objects in part of the area (+33$^{\rm o}<\delta<$+45$^{\rm o}$) of the First Byurakan Survey (FBS) which was aimed at the detection of blue stellar objects (Abrahamian \& Mickaelian \cite{abrahamian}, and references therein). To check the completeness of this sample we cross-correlated the sources in the ROSAT RASS X-ray survey (Voges et al. \cite{voges}) with the objects in the USNO-A2.0 catalog (Monet et al. \cite{monet}) brighter than O(USNO) = 16.0 and found a few bright additional QSO candidates. One of them, near \object{RX~J1643.7+3402}, is a bright starlike object at a galactic latitude of +40\fdg2, located 82\arcsec\ to the west of the bright F2 star \object{HD~151087} (see Fig. \ref{figureA0}). Its J2000 coordinates, measured on the Digitized Sky Survey, are: $\alpha$ = 16$^{\rm h}$ 43$^{\rm m}$ 49\fs71 and $\delta$ = 34$^{\rm o}$ 02\arcmin\ 39\farcs9; it is located 7\farcs2 from the X-ray position. This identification was first proposed by Bade et al. (\cite{bade}) and later by Rutledge et al. (\cite{rutledge}). The observations reported in this paper show it to be a cataclysmic variable.

\section{ Observations}
\subsection{ Spectroscopic observations}
On May 23, 2000, three spectra (10 min exposure each) were obtained with the ByuFOSC-2 focal reducer (Movsessian et al. \cite{movsessian}) attached to the prime focus of the Byurakan Observatory 2.6-m telescope; the ``green" grism (4200--6900 \AA) was used giving a dispersion of 2.7 \AA\ pix$^{-1}$. On May 25, another spectrum (15 min exposure) was obtained with the ``red" grism (5400--7600 \AA) giving a dispersion of 2.1 \AA\ pix$^{-1}$. The detector was a 1060$\times$514, 24$\times$24 $\mu$m pixel Thomson CCD.  The spectra were flux calibrated using the standard star \object{BD+28\degr4211} (Stone \cite{stone}). The continuum is very blue; several emission lines are present, the strongest being H$\alpha$, H$\beta$, He II $\lambda$4686, He I $\lambda$6678 and the ``Bowen blend" emission feature at $\sim$ 4645 \AA. \\

\begin{figure*}[t]
\resizebox{14.cm}{!}{\includegraphics{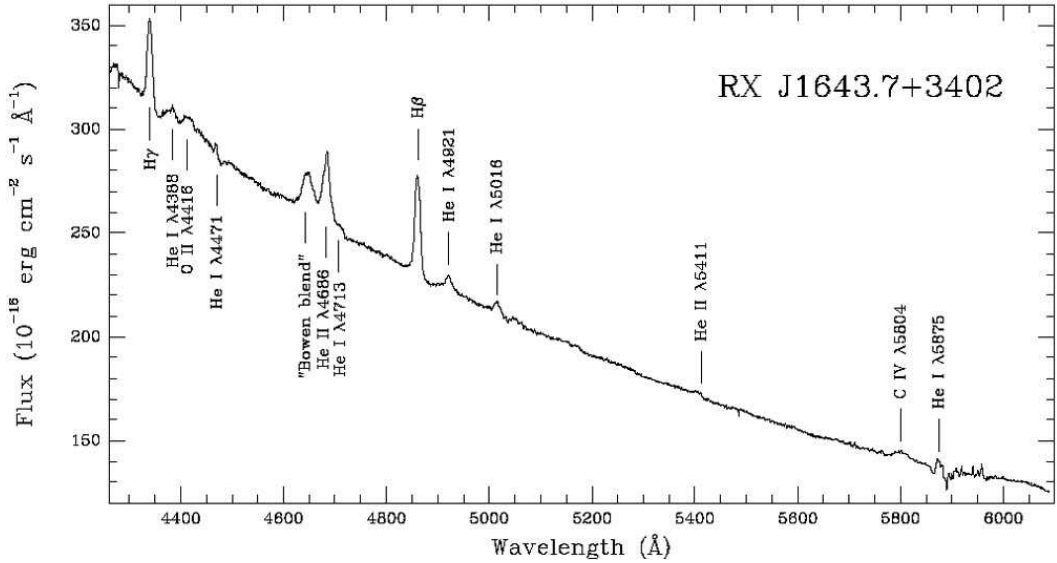}}
\hfill
\raisebox{5cm}{
\parbox[c]{3.8cm}{
\caption{\label{sp_polar}
Mean spectrum of RX~J1643.7+3402, obtained at the OHP 1.93-m telescope with the CARELEC spectrograph, showing the blue continuum, approximately $F_{\lambda} \propto \lambda^{-2.7}$, and the main emission lines discussed in the text (the total exposure time is 387 min).
}}}
\end{figure*}

During the period 22-28 September 2000 more detailed spectroscopic observations were carried out with the  CARELEC spectrograph (Lema\^{\i}tre et al. \cite{lemaitre}) attached to the Cassegrain focus of the OHP 1.93-m telescope. We used a 600 l\,mm$^{-1}$ grating resulting in a dispersion of 66 \AA\ mm$^{-1}$. The detector was a 1024$\times$2048, 13.5$\times$13.5 $\mu$m pixel EEV 42-20 CCD. Exposure times were generally 10 min. Seven columns of the CCD ($\sim$3\arcsec) were extracted. The slit width was 2\farcs1, corresponding to a projected slit width on the detector of 52 $\mu$m {\it i.e.} 3.8 pixels. The resolution, as measured on the night sky lines, was $\sim$ 3.4 \AA\ FWHM. The wavelength range was 4280--6100 \AA. The spectra were wavelength calibrated using an Ar lamp and flux calibrated using the standard stars \object{Feige\,15}, \object{BD+28\degr4211} and \object{EG\,247}, taken from Oke (\cite{oke}) and Stone (\cite{stone}).  To obtain accurate velocity measurements, we applied wavelength corrections by measuring the [O I] $\lambda$5577 night sky line on each spectrum (the corrections were in the range $-$1.3,$-$0.4 \AA).

The profiles of all emission lines are complex and variable. Some H$\beta$ profiles have a flat top, others display a shoulder while a few show a clear central absorption; we derived radial velocities using our own software, as described in Gon\c{c}alves et al. (\cite{goncalves}), by fitting each profile with a main emission component and a minor unresolved Gaussian absorption (see Section 5). This procedure derives velocity information essentially from the line flanks. The radial velocities for the main emission component vary between $-$205 and +125 km s$^{-1}$.

\begin{table}[h]
\caption{\label {EW}Equivalent widths of the main emission lines as measured 
on the integrated spectrum}
\begin{center}
\begin{tabular}{ll}
\hline
Line & EW (\AA) \\
\hline
 H$\gamma$           & 1.6  \\
 Bowen Blend         & 2.0  \\
 He II $\lambda$4686 & 2.5  \\
 He I $\lambda$4713  & 0.12 \\
 H$\beta$            & 2.8  \\
 He I $\lambda$4921  & 0.33 \\
 He I $\lambda$5016  & 0.32 \\
 C IV $\lambda$5804  & 0.62 \\
\hline
\end{tabular}
\end{center}
\end{table}
\normalsize

We coadded all our spectra (see Fig. \ref{sp_polar}). The resulting spectrum shows, in addition to the lines seen on the low resolution BAO spectra, several He~I lines (4388, 4471, 4713, 4921, 5016 and 5875 \AA) and an additional He~II line (5411 \AA); there is also a weak, broad ($\sim$33 \AA\ FWHM) C IV $\lambda$5804 emission line (this is in fact a doublet at 5801.5 and 5812.1 \AA\ ). Table \ref{EW} gives the equivalent widths (EW) of the main emission lines measured on this spectrum.  Blends of atmospheric water vapor absorption lines are present between 5680 and 5800 \AA\ and between 5870 and 6000 \AA.

\subsection{ Photometric observations}

We obtained photometric observations of the RX~J1643.7+3402 field on nine nights (18-27 September 2000) using the OHP 1.2-m telescope and a CCD TK1024$\times$1024 camera. The pixel size is 24 $\mu$m which projects to 0\farcs7  on the sky. Most of the frames used for monitoring the brightness were 30 sec V-band exposures. The effective time resolution (exposure time + read-out time) was 113 sec on the first four nights, but on subsequent nights we used a single 512$\times$512 detector quadrant (6$\times$6 arc-min field of view), which resulted in a read-out time of 30 sec and an effective time resolution of 60 sec. The longest runs possible at this time of year were about 2.5 to 3 h long. Total or partial simultaneous coverage with the spectroscopic observations at the 1.93-m was achieved on six nights (22-27 September). Dome flat-fields were obtained using controlled daylight illumination. All necessary data reduction was done using the ESO-MIDAS software package.

\begin{figure}[h]
\resizebox{8.8cm}{!}{\includegraphics{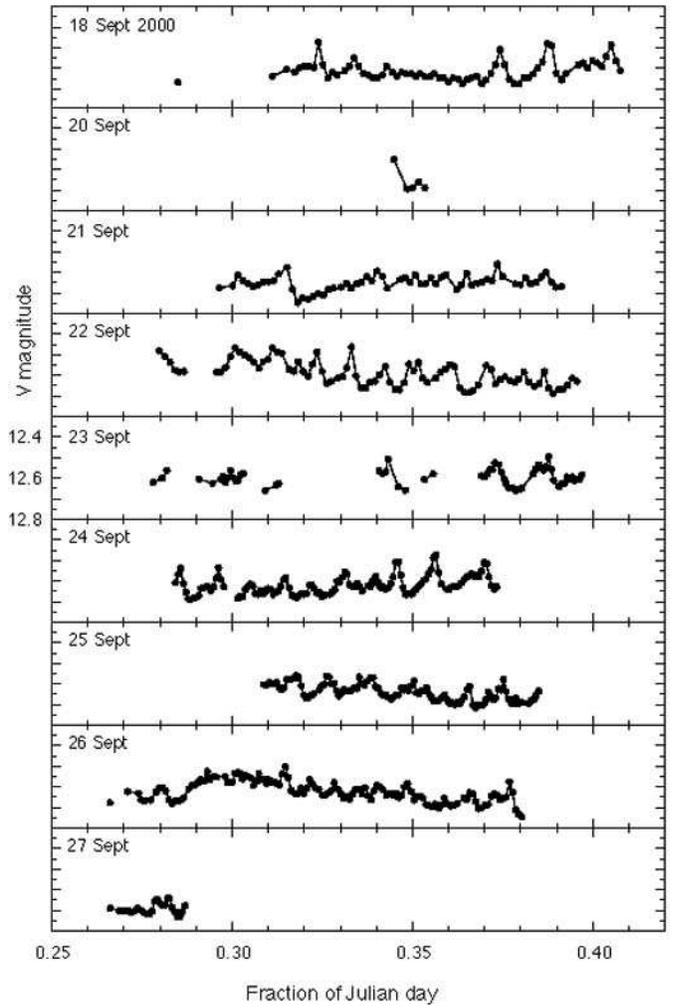}}
\caption{\label{figureA1}
CCD photometry of RX~J1643.7+3402 obtained with the 1.2-m OHP telescope in the V band on the dates shown. The magnitude scale is the same for all panels but is displayed only once. Note the strong (0.1 to 0.2 mag) rapid variations  with a 15 min  time scale which are present on most nights. The $\sim$ 0.1 mag periodic modulation at 9.25 c/d (2\fh595) reported in the text can be discerned best on the nights of September 18, 22, 24 and 26.}
\end{figure}
We derived differential magnitudes for the variable star and two nearby comparison stars using custom optimal-aperture photometry software. Sky values were derived for each star from concentric annuli designed to correct essentially for scattered light from the nearby bright star HD~151087. Results for our photometry runs are shown in Fig. \ref{figureA1} where the magnitudes, derived from 655 images, have been transformed into the standard V-band system. These light curves show that the object varied between V magnitude 12.5 and 12.7 during our September 2000 observing run. Superposed on a $\sim$ 0.1 mag slow modulation, there are rapid variations of 0.1 to 0.2 magnitude amplitude repeating with a time scale of $\sim$ 15 minutes.

Observations using the entire UBVR$_{\rm c}$I$_{\rm c}$ system filter set were secured on September 23, together with frames of standards in M13 (Forbes \& Dawson \cite{forbes}; Arp \& Johnson \cite{arp}) for photometric calibration. 
\begin{table}[h]
\caption{\label {table2}Magnitudes and colors of field stars shown in
the finding chart for RX~J1643.7+3402}
\begin{center}
\begin{tabular}{llcccc}
\hline
Name & V & U$-$B & B$-$V & V$-$R$_{\rm c}$ & R$_{\rm c}$$-$I$_{\rm c}$ \\
\hline
        Star 2  & 12.97 & $-$0.03 & +0.54 & +0.32 & +0.26 \\
        Star 3  & 12.16 & $+$0.94 & +1.04 & +0.61 & +0.44 \\
        Star 4  & 14.74 & $+$0.89 & +1.08 & +0.59 & +0.48 \\
        Star 5  & 16.30 & $-$0.21 & +0.43 & +0.34 & +0.29 \\
        Star 6  & 16.68 & $+$0.04 & +0.66 & +0.39 & +0.31 \\
        Star 7  & 16.43 & $+$1.18 & +1.56 & +1.02 & +1.01 \\
        Star 8  & 16.14 & $+$0.55 & +0.87 & +0.46 & +0.38 \\
\hline
\end{tabular}
\end{center}
\end{table}
\normalsize
Table \ref{table2} gives magnitudes and colors for several stars in the field including the two comparisons (stars 2 and 3, see Fig. \ref{figureA0}).  The calibration errors lie within $\pm$ 0.02 and $\pm$ 0.05 mag, depending on the magnitude.  Two different measurements of RX~J1643.7+3402 were obtained for each color index, except U$-$B. 
\begin{table}[h]
\caption{\label {table3}Magnitudes and colors of RX~J1643.7+3402 at two different times on 23 September 2000}
\begin{center}
\begin{tabular}{lcccc}
\hline
 V & U$-$B & B$-$V & V$-$R$_{\rm c}$ & R$_{\rm c}$$-$I$_{\rm c}$ \\
\hline
         12.58 & $-$0.95 & $-$0.07 & +0.02 & $-$0.01 \\
      	 12.65 &         & $-$0.04 & +0.04 & +0.02 \\
        \hline
\end{tabular}
\end{center}
\end{table}
\normalsize
The results are given in Table \ref{table3}. The rapid variability seen in the data of Fig. \ref{figureA1} may have slightly affected the derived colors since  measurements in any pair of filters were taken one minute apart (5 minutes in the case of the U filter). 

\section{Long-term variability}

We have investigated the long-term variability of this object by comparing the magnitudes derived from several Palomar Schmidt surveys. In the Guide Star Catalog GSC-ACT (Lasker et al. \cite{lasker}), which is based on the `Quick-V' Survey (the epoch for the relevant plate is 1982.6), the GSC V magnitudes for stars 2, 3 and 4 are 0.56 mag too bright with respect to the values in Table \ref{table2} and, correcting the catalog value for the variable, we derive V = 12.69, well within the range seen in Fig. \ref{figureA1}.

For the POSS-I survey blue and red plates, the USNO A2 catalog (Monet et al. \cite{monet}) gives O = 15.2 and E = 13.2. While the blue magnitude is very much fainter than our values (by 3 mag.), the Minnesota APS catalog (Pennington et al. \cite{pennington}), also based on the POSS-I survey, does not list either the variable or star 2 due to their proximity to the nearby bright star HD~151087. Thus we consider the USNO magnitudes for the variable to be unreliable (this is confirmed by fact that the USNO blue magnitude for star 2 is also too faint (by 1.3 mag), while the values for stars 3 to 8 differ little (by $-$0.07 mag) from our B magnitudes). In order to remedy this situation, we measured magnitudes for the variable and the stars of Table \ref{table2} on the Digital Sky Survey (DSS) POSS-I Red image (Epoch 1954.4) using our aperture photometry software and obtain R$_{\rm c}$ = 12.7 for the variable,  consistent with our CCD photometry.

For the POSS-II survey, we measured magnitudes for the same stars using the DSS blue (epoch 1989.5) and red (epoch 1993.3) images and obtain B = 13.2  and R$_{\rm c}$ = 12.7 for the variable. While the red magnitude is again in good agreement, the object appears fainter on the blue image by 0.6 mag. 

We conclude that the brightness of RX~J1643.7+3402 in the past does not appear to have undergone large variations relative to the values measured in September 2000.

\section{ Period search}

\subsection{Radial velocities}

We analyzed the H$\beta$ emission line velocities for periodic behavior using Discrete Fourier Transform (DFT) and Analysis-of-Variance (AoV) algorithms (as implemented in the MIDAS Time Series Analysis package). The short duration of the night runs (an average of 2\fh5), limited as it is in late September by hour-angle constraints, inevitably introduces a window function with a large number of 1 c/d aliases for any existing period.

\begin{figure}[h]
\resizebox{8.8cm}{!}{\includegraphics{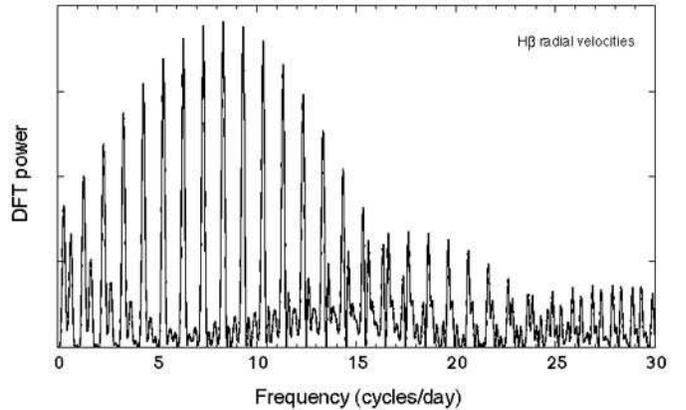}} 
\caption{\label{figureA2} 
The Discrete Fourier Transform (DFT) periodogram computed using 36 individual H$\beta$ emission line radial velocities is displayed here for the frequency interval 0 to 30 c/d. The strongest peaks are at 8.32 $\pm$1 c/d. The effects of the window function are apparent as aliases separated by 1 c/d.}
\end{figure}

\begin{figure}[h]
\resizebox{8.8cm}{!}{\includegraphics{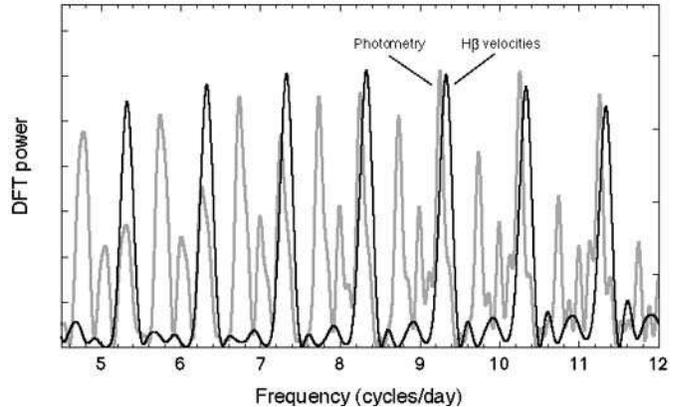}} 
\caption{\label{figureA3}
DFT periodograms in the frequency interval 4.5 to 12 c/d for both the H$\beta$ emission line radial velocities (dark curve) and the photometry (grey curve). The two curves have been normalized so that the strongest peaks in each appear of equal strength. The peaks  in the velocity and the photometry periodograms appear at different frequencies.  }
\end{figure}

The raw DFT spectrum, computed using a frequency step of 0.01 c/d, is shown in Fig. \ref{figureA2} for the interval 0-30 c/d. A more detailed plot is shown in Fig. \ref{figureA3} (thick dark curve) for the interval 4.5 to 12 c/d (periods from 5.3 to 2 h).  The strongest peaks are at 7.32, 8.32 and 9.32 c/d. Removing the effects of the window function using the CLEAN algorithm (Roberts et al. \cite{roberts}) produces a peak either at 8.32 (2\fh885) or at  9.32 c/d (2\fh575), depending on the values used for the gain (the amount by which the window function is subtracted from the raw DFT at each iteration), number of iterations, and frequency step.

\begin{figure}[h]
\resizebox{8.8cm}{!}{\includegraphics{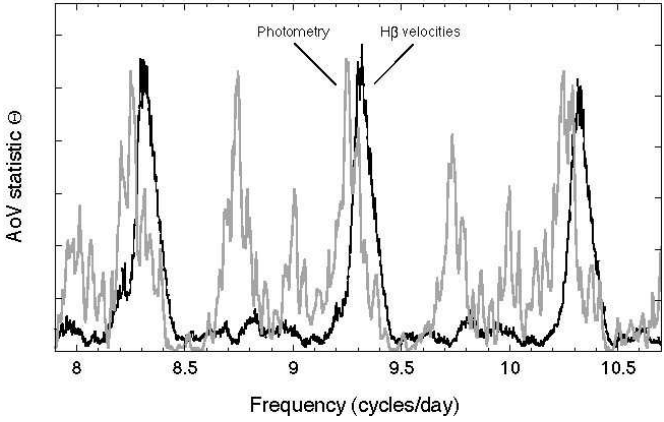}}
\caption{\label{figureA4}
AoV periodograms in the interval 7.9 to 10.7 c/d for both the H$\beta$ radial  velocities (dark curve) and the photometry (grey curve). The two curves have been normalized so that the strongest peaks in each appear of comparable strength. The peaks in the velocity and the photometry periodograms appear at different frequencies. }
\end{figure}

We also computed AoV periodograms (Schwarzenberg-Czerny \cite{schwarz89}) for the H$\beta$ emission line radial velocity data using different numbers of phase bins and various amounts of smoothing. In Fig. \ref{figureA4} we show the result for 8 bins and 3 overlapping bin covers (thick dark curve). The three strong peaks are at 8.32, 9.32 and 10.32 c/d. The peak at 9.32 c/d (2\fh575) appears slightly stronger here, but the high-frequency noise present in the AoV periodogram, which arises from the small number of data points which move around from one phase bin to the next as the period changes, may mask the true strength of the peaks.

\begin{figure}[h]
\resizebox{8.8cm}{!}{\includegraphics{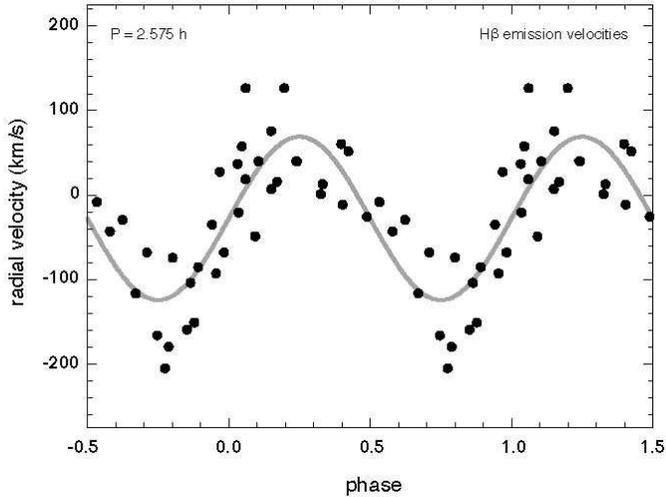}}
\caption{\label{figureA5}
H$\beta$ emission line radial velocities folded using the 9.32 c/d (2\fh575) period. The epoch for phase 0.0 is HJD 2451806.0324. The semi-amplitude is K = 97 $\pm$ 12 km s$^{-1}$ and the heliocentric systemic velocity is $\gamma$ = $-$28 $\pm$ 9 km s$^{-1}$. The curve for the other possible spectroscopic period, 8.32 c/d (2\fh885), is nearly identical. }
\end{figure}

The results of our analysis are ambiguous. The existing radial velocity data appears insufficient to decide which of the two periods, 8.32 c/d (2\fh885) or 9.32 c/d (2\fh575), is the real orbital period of RX~J1643.7+3402.  A least-squares sine fit to the velocities gives an estimate for the errors on the periods of $\pm$ 0.013 c/d ($\pm$ 0\fh005).

 The radial velocities are plotted in Fig. \ref{figureA5} as a function of phase (for the 2\fh575 period; the plot for the 2\fh885 period is nearly identical). The semi-amplitude is K = 97 $\pm$ 12 km s$^{-1}$. In the absence of any clear specific marker for orbital orientation, such as an eclipse, the epoch for phase 0 (HJD 2451806.0324 for the 2\fh575 period, HJD 2451806.1213 for the 2\fh885 period) has been defined  as the time when the radial velocities change sign from negative to positive with respect to the heliocentric systemic velocity, which is $\gamma$ = $-$28 $\pm$ 9 km s$^{-1}$. 
Using this convention, and if the emission-line velocities reflect the motion of the white dwarf, which they may not, inferior conjunction of the white dwarf would take place at phase $\phi = 0.0$.\\

\subsection{Light curves}

In order to search for the two possible periods detected in the radial-velocity data, and to look for a short-period modulation, we analyzed the photometric data with the DFT+CLEAN and AoV algorithms. We excluded from the analysis the very short nights of September 20 and 27, the night of September 23 which exhibits large gaps and the night of September 21 which, apart from a sharp discontinuity, appears flat and featureless. The selected set contains the data for September 18, 22, 24, 25 and 26, 502 images in all.

The resulting DFT periodogram was computed using a frequency step of 0.005 c/d and is shown in Fig. \ref{figureA3} (gray curve) together with the periodogram of the radial velocity data.  It is also severely dominated by the window function but appears more complex (with two interleaved sets of peaks). The strongest peak appears at 9.25 c/d with an equally strong peak at 10.25 c/d. The results of CLEANing the DFT spectrum show only a strong peak near 9.25 c/d (2\fh595), with a second weaker peak near 2.74 c/d (8\fh76) with one-half the power.

The results of an AoV analysis of the photometry are also shown in Fig. \ref{figureA4} (grey curve), together with the periodogram of the radial-velocity data. The 9.25 c/d peak appears stronger than the peak at 10.25 c/d, and also stronger than the peak at 8.25 c/d. The peaks in Figs. \ref{figureA3} and \ref{figureA4} derived from the photometry analysis have frequencies (8.25 and 9.25 c/d) that are similar to, but different from, those found from the analysis of the radial velocity data (8.32 and 9.32 c/d). The difference is small (0.8\%) but significant.

\begin{figure}[t]
\resizebox{8.8cm}{!}{\includegraphics{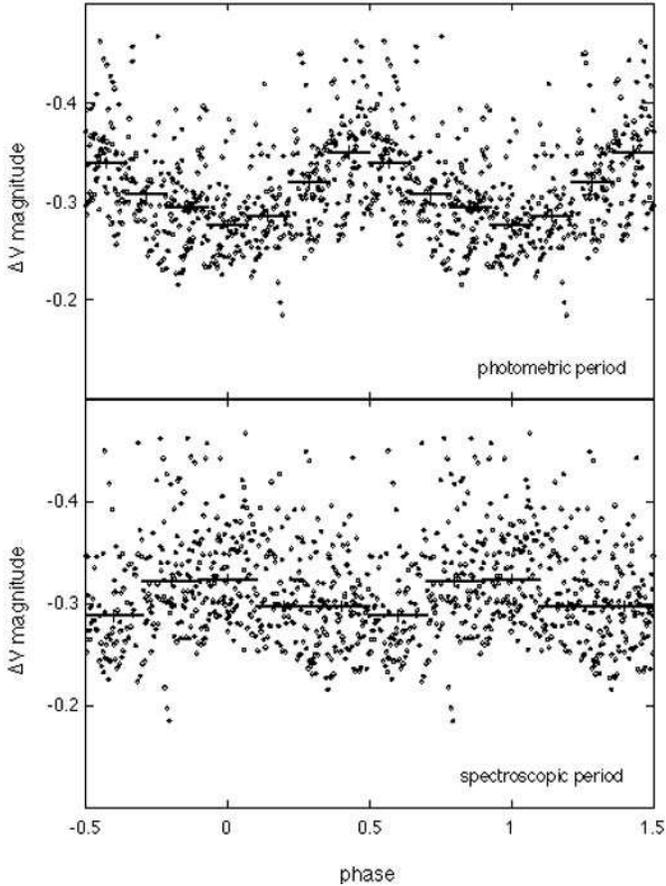}}
\caption{\label{figureA6}
CCD V-band differential light curves constructed by folding the individual measurements for a selected subset of nights (September 18, 22, 24, 25 and 26) using one set of possible periods (a photometric period at 9.25 c/d (or 2\fh595)  and a spectroscopic period at 9.32 c/d (or 2\fh575)). The curves for the other possible periods are similar. No corrections for the small night-to-night changes in mean level have been applied here. Epoch for phase 0 is HJD 2451806.036. The magnitudes are expressed with respect to star 2. The data are also shown averaged into 7 and 5 bins, respectively, with the corresponding errors on the mean values. The  peak-to-peak amplitude in the upper panel is $\sim$ 0.1 mag.}
\end{figure}

When the differential photometry data for individual nights are folded using one of the spectroscopic periods, a modulation is seen which clearly shifts forward in phase from night to night, but when the data are folded using the closest photometric period, the modulations do not shift and are in phase with each other. The results for the selected subset of nights (September 18, 22, 24, 25 and 26) are shown in Fig. \ref{figureA6} folded using one of the possible photometric periods (the others yield similar curves). Although small (1 to 2\%) night-to-night changes in mean level are present, no corrections have been applied to the data plotted in this figure. There is a significant average modulation of $\sim$ 0.1 mag  peak-to-peak amplitude at 9.25 c/d (2\fh595) but little or none is seen at 9.32 c/d (2\fh575), the value found from the radial velocities. The modulation can be discerned best in the light curves of Fig. \ref{figureA1} during the nights of September 18, 22, 24 and 26.

In summary, one likely interpretation of our results is that the radial velocities are modulated with a 9.32 c/d (2\fh575) period while the photometric data show a sligthly longer (0.8\%) period of 9.25 c/d (2\fh595). However, an alternative interpretation where the radial-velocity data are modulated with a 8.32 c/d (2\fh885) period while the photometric data display a noticeably shorter (10\%) period of 9.25 c/d (2\fh595) is also possible. Note that the other photometric periods (at $\pm$ 1 c/d) are not entirely excluded. The implications of these two different interpretations are discussed later.

\subsection{Rapid variability}

None of our nights was sufficiently long to do a reliable time analysis within a single night for periods around 15 minutes and we looked instead for a coherent period in our data by grouping the longest nights with uninterrupted coverage which clearly displayed the rapid variability (September 18, 22, 24, 25, 26) and subjected them to an AoV analysis (within the 50-150 c/d interval) using three phase bins and no smoothing. There is no single peak which stands out in the resulting AoV periodogram, but the strongest signals are located between 80-85 c/d (18-17 min) and around 100 c/d (14 min), confirming the visual impression of a time scale around 15 min. The CLEANed DFT spectrum agrees with these results showing power principally near 84.2 c/d and at 101.8 c/d, with the latter peak being the strongest.

\section{Spectroscopic characteristics}

\subsection{Balmer and He~I lines}

\begin{figure}[t]
\resizebox{8.8cm}{!}{\includegraphics{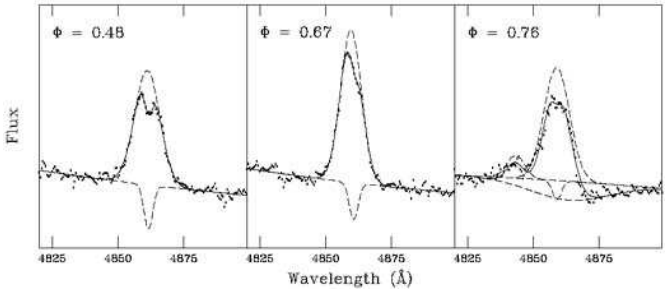}}
\hfill
\caption{\label{polar_hb}
Three examples of the fit of the H$\beta$ feature with Gaussian components. The leftmost panel shows a profile with a clear narrow absorption component. The middle panel show the case of a profile displaying a ``shoulder," accounted for by a red-shifted absorption component. The right panel shows a case with a ``flat-topped" profile where both blue emission and red absorption components were needed for a good fit. The spectra have been normalized through division by the continuum.
}
\end{figure}

The profiles of all emission lines are complex and variable. A few of the H$\beta$ profiles show a strong, narrow absorption; but most have a flat top or a shoulder, suggesting the presence of weak absorption in the line core. In addition, some of the profiles show weak blue emission, and/or red absorption. We fitted each of the H$\beta$ profiles with several Gaussian components
(see Fig. \ref{polar_hb}): (1) an emission component, the ``main" component, (2) a narrow absorption component (assumed to be unresolved and thus assigned a fixed width of 4.0 \AA\ FWHM), (3) a blue emission component and (4) a broad red absorption component. Due to their location in the line wings and to their relative weakness the two latter components were constrained, the first in width (fixed at 8.0 \AA\ FWHM) and the second in width (fixed at 35 \AA\ FWHM) and in velocity (fixed at +135 km s$^{-1}$). 

We find that the main component has a variable FWHM (8.0--15.0 \AA), but half of the values are in the range 10.6--11.8 \AA. The analysis of H$\gamma$ and He~I $\lambda$5875 gives similar results; the velocities of the main component of the three lines are well correlated. The H$\beta$ and H$\gamma$ FWHMs are also correlated, suggesting that the observed variability is real and not the result of measurement errors. We also note that the presence, around phases $\phi \approx$ 0.8--0.2, of a blue emission and of a red absorption in  H$\beta$ are well correlated in the sense that, when a red absorption is present, a blue emission is always present; moreover, we never observe a red emission nor a blue absorption. 

The velocities of the narrow Gaussian absorption component fitted to the H$\beta$ profiles (see Fig. \ref{polar_hb}) were measured to have a semi-amplitude K = 107 $\pm$ 19 km s$^{-1}$ and an average velocity $\gamma$ = +6 $\pm$ 13 km s$^{-1}$. They follow closely the main H$\beta$ emission component as there is no significant phase shift (0.035 $\pm$ 0.034). The intensities of the main Gaussian component fitted to the data are seen to vary in phase with a minimum in the phase range 0.9--0.3 and a maximum near 0.6-0.8. The high-velocity blue emission was measured on 15 spectra and its velocities give a semi-amplitude K = 280 $\pm$ 110 km s$^{-1}$ and an average velocity $\gamma$ = $-$890 $\pm$ 76 km s$^{-1}$. The phase of maximum blue shift is 0.95 $\pm$ 0.06. There is one spectrum where the velocity was measured to be $-$1880 km s$^{-1}$, but others lie within the interval $-650$ to $-1200$ km s$^{-1}$. The curve of the H$\beta$ high-velocity component shows a phase lag of 0.2 with respect to the main emission. 
\\

\begin{figure}[t]
\resizebox{8.8cm}{!}{\includegraphics{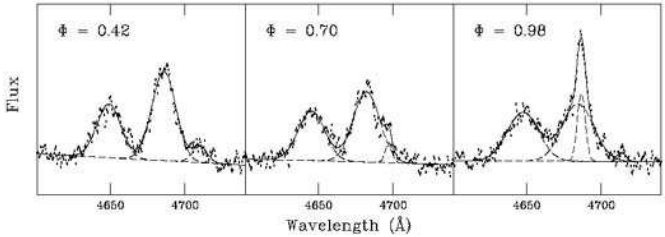}}
\hfill
\caption{\label{heII}
Three examples of the fit of the He II $\lambda$4686 line and the ``Bowen blend" with Gaussian components. The spectra have been normalized by dividing them by the continuum. The ``Bowen blend" has been fitted by a combined C III feature at $\lambda$4649; the velocity and width of the C III line has been forced to be equal to that of the broad He II line. In the left panel there is no He II narrow component, but the He I $\lambda$4713 line is visible. In the middle panel there is a He II red-shifted narrow component, while in the right panel the He II line has a strong narrow component.
} 
\end{figure}

\subsection{``Bowen blend" and He~II $\lambda$4686}
The ``Bowen blend" is often observed in binary X-ray sources (McClintock et al. \cite{mcclintock}); in \object{Sco~X-1}, its constituents are N~III $\lambda\lambda$4634, 4641 and 4642 and C~III $\lambda\lambda$4647, 4651 and 4652; the N III lines account for about two-thirds of the total blend intensity; the theoretical ratio I($\lambda$4634)/I($\lambda\lambda$4641,4642) is 0.71 (Schachter et al. \cite{schachter}). The N~III lines are produced by fluorescence, the C~III lines are not (Bowen \cite{bowen}; Ferland \cite{ferland}).

We analysed the spectral region around He II $\lambda$4686, which contains the ``Bowen blend," by using six Gaussian profiles: the first two, intended to represent the N~III fluorescent lines, were centered at 4634.2 and 4641.0 \AA\ respectively; they were forced to have the same width and velocity; the flux of the first was assumed to be half that of the second; the third profile, centered at 4649.0 \AA, corresponded to the C~III triplet; two more profiles were judged necessary to fit reasonably the He II line itself; the sixth component corresponds to He~I $\lambda$4713.1; we fixed its width to 12 \AA\ (FWHM), the average width found for the Balmer main emission components. The fit yielded a very weak, unsignificant N~III system in emission, suggesting that, in RX~J1643.7+3402, the C~III triplet accounts for most of the ``Bowen blend". The average FWHM of both the C~III triplet at $\lambda$4650 and the C~IV doublet at $\sim \lambda$5804 is about 1500 km s$^{-1}$.

The Gaussian fits to the He II emission line required two different components, a ``broad" component and a ``narrow" component, examples of which are illustrated in Fig. \ref{heII}. The velocities of the ``broad" component vary between 0 and $-$350 km s$^{-1}$. The semi-amplitude is K$_b$ = 186 $\pm$ 25 km s$^{-1}$ and the mean velocity is $\gamma_b$ = $-$197 $\pm$ 18 km s$^{-1}$. The phase lag  with respect to the velocities of the main H$\beta$ emission component is very small (0.054 $\pm$ 0.030) and may in fact be negligeable. The ``narrow" He~II component shows a curve similar to the ``broad" component but is shifted in velocity by $\sim$ +300 km s$^{-1}$ (see Fig. \ref{heii-3}); this suggests that these two components probably describe a complex single profile. Two individual spectra show a particularly striking red-shifted ``narrow" component (see Figs. \ref{heII} and \ref{heii-3}) near phase 0.75. Phase coverage for the ``narrow" component is uneven since none is seen in some spectra (phase 0.42) while in others its absence may in part be due to a poor S/N ratio.

\begin{figure*}[t]
\resizebox{18.0cm}{!}{\includegraphics{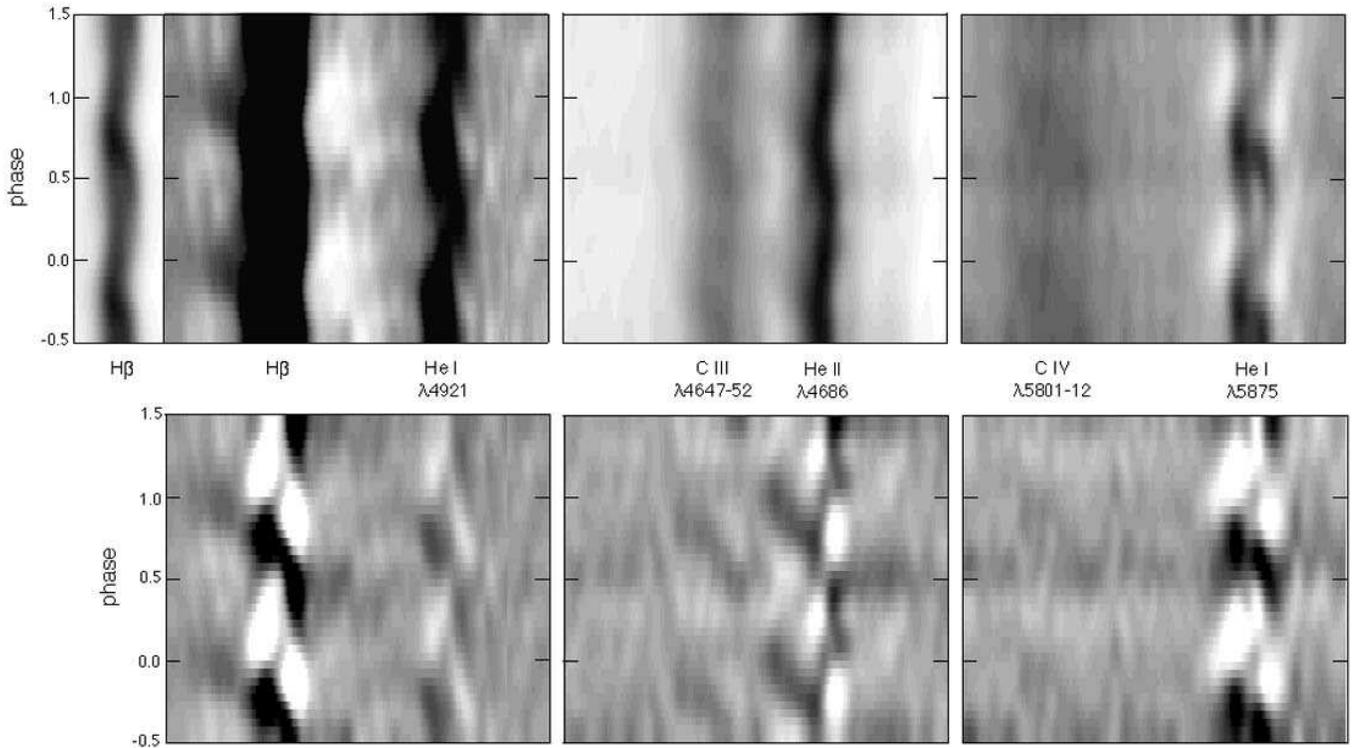}}
\caption{\label{trailed}
{\bf (Top panel)} ``Trailed" spectra built by folding all normalized spectra into ten phase bins assuming a period of 2\fh575. The spectra were then smoothed as described in the text. The result is displayed using 20 phase bins. All narrow lines (H$\beta$, He~I and  He~II) have a sinusoidal shape; the H$\beta$ line is shown twice with different contrasts; the He~I $\lambda$5875 line clearly shows a central absorption component which is most conspicuous at phases $\sim$0.3--0.6. The Na D absorption lines to the red of He~I $\lambda$5875 did not display any detectable phase shift and have been numerically removed to reveal the red He I absorption components. {\bf (Bottom panel)} The residuals obtained by substracting the phase-averaged profile from the ``trailed" spectra shown in the top panel. This enhances the relative changes in the emission lines as a function of phase. Complex low-velocity modulations, having a distinct ``braided" appearance, are seen in the Balmer lines, the He~I lines and the He~II line. Note the high-velocity blue-shifted He~II S-wave visible for one-half of the cycle. The wavelength interval shown for each panel is 140 \AA. 
}
\end{figure*}

\subsection{``Trailed" spectra}
We constructed ''trailed" spectra by folding all the normalized spectra into 10 phase bins using the 2\fh575 period. The spectra were then smoothed in wavelength using a 3-point running average and in phase using a $10\times1$ Gaussian filter. The result is displayed using 20 phase bins. Fig. \ref{trailed}(top panel) shows three sections of this spectrum (each panel is 140 \AA\ wide) including respectively H$\beta$ and He I $\lambda$4921, the ``Bowen blend" and He~II $\lambda$4686, and C~IV $\lambda$5804 and He~I $\lambda$5875; all narrow lines (H$\beta$, He~I and He~II) show a clear S-shape. The He~I $\lambda$5875 line has three main components: an emission component behaving like H$\beta$, a broad absorption component and a narrow (unresolved?) absorption component visible mainly at phases $\sim$~0.3-0.6. The H$\beta$ line (shown twice with different contrasts) has a strong emission core with a narrow absorption component (barely visible on the figure) which is most conspicuous at the same phases as in the He~I line; in addition, a very weak broad blue emission appears during about half the period, while a broad red absorption is detected simultaneously. The He~II line has a narrow core similar to the H$\beta$ core, but without any trace of the narrow absorption component; in addition, there is a broad emission component.

The bottom panel of Fig. \ref{trailed} shows the result of subtracting the phase-averaged profile from the ``trailed" spectra shown in the top panel. This enhances the relative changes in the emission lines. Complex low-velocity  modulations, which have a distinct ``braided" appearance, are seen in the Balmer lines, the He~I lines and the He~II line. The modulations in the Balmer and He~I lines are similar, although the He~I $\lambda$5875 line profile is complicated by the stronger central absorption. In the case of the He~II $\lambda$4686 line there is an additional blue-shifted high-velocity S-wave which is clearly visible for roughly one-half of the orbital cycle, from phases 0.6 to 0.2, roughly opposite to the time when the central absorption is strongest in the Balmer and He~I lines. This blue-shifted He~II emission S-wave  also lags in phase by 0.2 with respect to the ``main" profile, showing roughly the same velocity curve as the blue-shifted H$\beta$ emission component. 

\begin{figure}[h]
\resizebox{8.8cm}{!}{\includegraphics{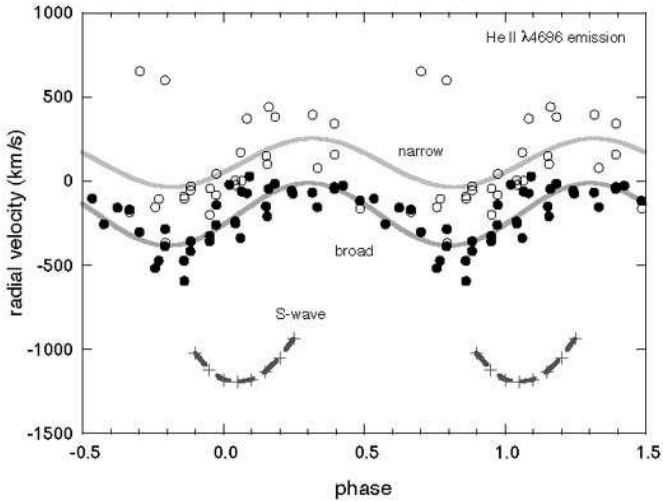}}
\caption{\label{heii-3}
He~II $\lambda$4686 radial velocities are shown for the two components of the ``main" emission (``broad" and ``narrow") and for the blue S-wave.  The period and epoch used are the same as for Fig. \ref{figureA5}. For the ``main" components we show the least-squares sine fits. The semi-amplitudes are K$_{b}$ = 186 $\pm$ 25 km s$^{-1}$ and K$_{n}$ = 144 $\pm$ 65 km s$^{-1}$ and the average velocities are $\gamma_{b}$ = $-$197 $\pm$ 18 km s$^{-1}$ and $\gamma_{n}$ = +110 $\pm$ 52 km s$^{-1}$. For the S-wave component the velocities were measured on the residual ``trailed" spectra shown in Fig. \ref{trailed} (bottom panel).}
\end{figure}

There is a weakly visible S-wave both in the ``Bowen blend" and an even weaker one in the C~IV blend which is seen in the phase range from 0.4 to 0.9, both moving in sympathy with the blue-shifted He~II S-wave.  The He~I $\lambda$5875 line displays blue-shifted absorption when both the H$\beta$ and the He II emission show blue-shifted emission. 
In Fig. \ref{heii-3} we have plotted the radial velocities of the He~II $\lambda$4686 S-wave, as measured on the residual ``trailed" spectra, together with the velocities of the two components (``broad" and ``narrow") of the ``main" emission discussed earlier.

\section{ Discussion}

\subsection{ What kind of system ?}

The blue continuum, the relatively weak, high excitation emission lines and the relative long-term stability in brightness suggest that RX J1643.7+3402 belongs to the novalike class of cataclysmic variables (CVs). These objects are believed to transfer mass at a rate high enough so that they are stuck permanently in outburst, thus preventing the instabilities associated with dwarf novae. The spectrum shown in Fig. \ref{sp_polar} resembles those of V795 Her (Casares et al. 1996; Dickinson et al. 1997), DW UMa (Shafter et al. \cite{shafter88}), LX Ser (Young et al. \cite{young}) and V442 Oph (Hoard et al. \cite{hoard2000}) and looks very similar to that of V533 Her (Thorstensen \& Taylor \cite{thorstensen}). All these systems belong to the SW Sex sub-group of novalike CVs. The distinctive characteristics of SW Sex stars are (1) single-peaked emission lines, particularly He II $\lambda$4686, instead of the double-peaked lines typical of a Keplerian disk, (2) gross asymmetries in the emission lines from the disk, so that they do not reflect the motion of the white dwarf, (3) about two-thirds of the systems (11 out of 16) display eclipses of the accretion disk by the low-mass late-type companion, (4) peculiar absorption features in most lines at a phase opposite the eclipse, near maximum blueshift of the Balmer emission lines, (5) frequently, high-velocity S-waves in the Balmer emission lines, and (6) orbital periods grouped mostly in the 3--4-h range, but with two objects in the period ``gap" and one having an 8-h period. The two SW Sex stars with the shortest periods, V442 Oph (P = 2\fh98) and V795 Her (P = 2\fh60) are non-eclipsing systems (Hoard et al. \cite{hoard2000}; Casares et al. \cite{casares}). 

The ``trailed" spectra of Fig. \ref{trailed} show phase-dependent absorption reminiscent of that characterizing the SW Sex stars (see Thorstensen et al. \cite{thorstensena}). However, the absorption we see is strongest in the phase range 0.3-0.6, near the positive-to-negative cross-over point for the H$\beta$ emission-line velocities, whereas in SW Sex stars the absorption is normally strongest near the phase of maximum approach velocity, approximately one-half cycle away from the eclipses seen in most of these systems. As no eclipses have been detected in our photometry we cannot determine the relation between the radial velocities and the binary system orientation, but the relative difference in the times of appearance of the strongest absorption in SW Sex type stars and in our object, roughly one-quarter of an orbital cycle, is very large and does not depend on any particular phase convention. The UBV colors measured for RX~J1643.7+3402 are very similar to those reported for the non-eclipsing systems V442 Oph and V795 Her (see the compilation by Hoard \cite{hoard98}), perhaps a consequence of their accretion disks being viewed at lower inclinations.

There is currently no single agreed-upon explanation for the SW Sex phenomenon, but it appears that the accretion disks of these systems have a complicated structure. For example, the flared-out disk in DW UMa (Knigge et al. \cite{knigge})  masks parts of the disk surface at any given time and can thus explain the gross asymmetries and the single-peaked profiles of the emission lines (with an additional contribution from a disk wind) while substantial stream-disk overflow (Hellier \cite{hellier98}) may explain the absorption in the line cores, when the gas stream is seen projected onto the disk surface. The high-velocity S-wave emission features may be due to stream re-impact in an area near the disk center, where the Keplerian velocities are highest (Hellier \& Robinson \cite{hellierb}). The detailed interpretation of these S-waves is under debate. Casares et al. (\cite{casares}) consider the strong red and the weak blue wings in the V795 Her system as two separate components, perhaps related to accretion onto a magnetic white dwarf which is rotating synchronously, while Dickinson et al. (\cite{dickinson}) explain the two wings of that system as parts of a very broad emission S-wave with a sympathetically phased red-shifted superimposed absorption S-wave. In RX~J1643.7+3402 only the blue wing is seen in emission. The H$\beta$ profile shown in the right panel of Fig. \ref{polar_hb} is similar to those observed in V795 Her.

The suggestion by Horne (1999) that a disk-anchored magnetic propeller (a rapidly spinning magnetic white dwarf) may help  explain the SW Sex phenomenon has received some support with the recent discovery of variable circular polarization in LS Peg (Rodriguez-Gil et al. \cite{rodriguez}). This would suggest that SW Sex systems may just be intermediate polars with high accretion rates. 

RX~J1643.7+3402 shares several characteristics with SW Sex stars, such as  single-peaked instead of double-peaked emission lines,  asymmetric profiles,  high-velocity S-waves (although in this system we only see the blue part of this S-wave) and  absorption features in the line cores. It has, in particular, many properties in common with V795 Her, including similar spectroscopic period, rapid photometric behavior and superhump modulation. The 0.2 phase lag we find between the high-velocity S-wave and the main emission component for both He~II and H$\beta$ has also been found in the SW Sex system V1315~Aql (Hellier \cite{hellier96}). This phase lag has been interpreted as arising from the difference in the location, within the disk, of the two emission regions: the high velocity component coming from the inner disk (re-impact area ?) and the lower velocity component from a more external location (hot spot ?). The phasing of the absorption features in RX~J1643.7+3402, which are strongest when the main emission velocities change from recession to approach, is nevertheless different from that in SW Sex systems, where they appear close to the time of maximum approach velocity, suggesting a different geometry in the accretion stream disk-overflow.

\subsection {Optical modulation}
In the section on the period search we have seen that the exact value of the spectroscopic period of this system is not yet known and may be either 2\fh575 or 2\fh885, both within the period ``gap" (Shafter \cite{shafter1992}), which although sparsely populated, contains several novalike systems. The  photometric modulation detected on most nights (of $\sim$ 0.1 mag amplitude in V) at a probable period of 2\fh595  (but the $\pm$ 1 c/d aliases are not excluded), different from either of the spectroscopic periods, could be explained as a superhump modulation. In many dwarf novae, but also in novalike variables, photometric modulations are seen with a period different from the spectroscopic period (Taylor et al. \cite{taylor98}). This phenomenon is known as the superhump phenomenon and corresponds to the light variations of a precessing elongated accretion disk. In the case of dwarf-novae the superhumps appear only in outburst but, in novalike systems, the phenomenon is  persistent  and has been called ``permanent" superhumps. While most superhump periods are longer than the orbital period (``positive" superhumps), some appear at a shorter period (``negative" superhumps). It has been suggested that the former arise from the prograde motion of the line of apsides of an elongated disk, while the latter appear to imply a tilted, elongated disk displaying retrograde motion of the line of nodes (Patterson et al. \cite{patterson97}).  

Patterson (\cite{patterson99}) has reported that among a sample of 18 non-dwarf nova systems, 7 show positive superhumps, 4 show negative superhumps and 7 show both kinds. Essentially all novalike systems with P$_{orb} < 3$-h show positive superhumps. Only novalike systems show negative superhumps and, amongst these, four are SW Sex systems. Patterson also shows that the fractional period excess $\epsilon = (P_{sh} - P_{orb})/P_{orb}$ shows a correlation with orbital period, with negative superhumpers showing a period excess about one-half that for positive superhumpers. For the spectroscopic periods reported here, these correlations would predict $\epsilon \sim$ +0.07 for a positive superhump and $\epsilon \sim-$0.03 for a negative one. The values we apparently find, +0.008 or $-$0.10, are significantly different. Further observations will reveal the true orbital period of RX J1643.7+3402 and should allow a more detailed characterization of the optical modulation.

\subsection {Rapid variability}

While rapid variability is typical of most cataclysmic variables, the behavior we see in this object seems unusual, since the variations show a large amplitude (0.1 to 0.2 mag) and each peak in brightness seems well separated from its neighbor. Rapid variations of a similar character have been reported in the case of V795~Her (Zhang et al. \cite{zhang}). These variations have been variously interpreted in the literature as (1) oscillations of the disk boundary layer, (2) reprocessing of X-rays in ``blobs" of optically thick material orbiting within the accretion disk, and (3) accretion of ``blobs" onto the magnetic poles of a magnetized white dwarf. Longer observing runs are needed to obtain more accurate information on the characteristics of the rapid variations in RX J1643.7+3402.

\subsection{Distance and X-ray luminosity}

The average value for the EW of the H$\beta$ emission line for this object is  2.8 \AA. The correlation between the absolute visual magnitude and the H$\beta$ EW derived by Patterson (\cite{patterson84}) for cataclysmic variables  suggests that the absolute magnitude might be  brighter than M$_{\rm V}$ = +6. Assuming negligeable absorption, this yields a lower limit for the distance of 210 pc. For novalike systems Warner (\cite{warner}) discusses the correlation  between absolute visual magnitude and orbital period. For the period reported here this correlation yields M$_{\rm V}$ = +5 $\pm$ 1, consistent with the previous estimate. Since these values refer to a standard inclination angle of $\sim57^{\rm o}$, the correction in this case is probably small since no eclipses are detected. The entry in the ROSAT bright source catalogue gives a PSPC (0.1-2.4 keV) count rate of 0.05 $\pm$ 0.01 count s$^{-1}$ and a hardness ratio HR1 of 0.62 $\pm$ 0.22. Using the conversion factor given by Fleming et al. (\cite{fleming}) the X-ray flux is $7.1\times10^{-13}$ erg cm$^{-2}$ s$^{-1}$. We note that V795 Her has also been detected by ROSAT at approximately the same flux level (Rosen et al. \cite{rosen}). Assuming the distance derived above and no X-ray absorption this yields a very approximate 0.1-2.4 keV X-ray luminosity of $3.8\times10^{30}$ erg s$^{-1}$, which is near the lower end of the range observed for CVs (van Teeseling \& Verbunt \cite{teeseling2}, van Teeseling et al. \cite{teeseling}). The X-rays may be produced in the boundary layer between the white dwarf and the accretion disk but more X-ray data are needed to investigate the origin of the emission.

\section{Conclusion}

We have reported the discovery of a new bright (V$\sim$12.6) cataclysmic variable star identified with the ROSAT X-ray source RX~J1643.7+3402. Spectroscopic and photometric observations show it to be a novalike cataclysmic variable sharing several of the characteristics of the SW Sex sub-class of  novalike CVs. The spectroscopic period may be either 2\fh575 or 2\fh885, both within the period ``gap." A photometric modulation (of $\sim$ 0.1 mag amplitude in V) with a probable period of 2\fh595 is present on most nights and could be either a ``positive" or a ``negative" superhump modulation  (depending on the exact spectroscopic period), indicating the presence of a precessing accretion disk in this system. Rapid variations of 0.1 to 0.2 mag amplitude in V repeat with a time scale of 15 min. More observations are needed to resolve the spectroscopic period ambiguity, to study the optical modulation and refine its period and to fully characterize the rapid variations.

\begin {acknowledgements}

This work was supported by the JUMELAGE ``Astrophysique France-Arm\'enie," a bilateral program funded by the French Centre National de la Recherche Scientifique and the French Minist\`ere des Affaires Etrang\`eres. The Guide Star Catalog and the Digitized Sky Survey were produced at the Space Telescope Science Institute under U.S. Government grants. 

\end{acknowledgements}

\end{document}